\documentclass[a4paper,11pt]{article}
\usepackage{amsmath,amssymb,bm}
\usepackage[T1]{fontenc}
\usepackage{yfonts}
\usepackage{mathtools}
\usepackage{dsfont}
\usepackage[utf8]{inputenc}
\pdfoutput=1
\usepackage[english]{babel}
\usepackage{graphicx} 
\usepackage{booktabs} 
\usepackage{color}
\definecolor{aquamarine}{rgb}{0.5, 1.0, 0.83}
\definecolor{ao(english)}{rgb}{0.0, 0.5, 0.0}
\definecolor{armygreen}{rgb}{0.29, 0.33, 0.13}
\definecolor{awesome}{rgb}{1.0, 0.13, 0.32}
\definecolor{ballblue}{rgb}{0.13, 0.67, 0.8}
\definecolor{bittersweet}{rgb}{1.0, 0.44, 0.37}
\definecolor{blue}{rgb}{0.0, 0.0, 1.0}
\definecolor{brinkpink}{rgb}{0.98, 0.38, 0.5}
\definecolor{ballblue}{rgb}{0.13, 0.67, 0.8}
\definecolor{brightturquoise}{rgb}{0.03, 0.91, 0.87}
\definecolor{blue-green}{rgb}{0.0, 0.87, 0.87}
\definecolor{caribbeangreen}{rgb}{0.0, 0.8, 0.6}
\definecolor{cyan}{rgb}{0.0, 1.0, 1.0}
\definecolor{amber(sae/ece)}{rgb}{1.0, 0.49, 0.0}
\graphicspath{ {images/} }

\author{J\"{u}rg Fr\"{o}hlich\footnote{Theoretical Physics, ETH Zurich, Wolfgang-Pauli-Str.\,27, 8093 Zurich, 
Switzerland/ juerg@phys.ethz.ch}}

\title{Relativistic Quantum Theory}

\begin{document}

\maketitle

\begin{abstract}
The purpose of this paper is to sketch an approach towards a reconciliation of quantum theory	
with relativity theory. It will actually be argued that these two theories ultimately rely on one another. 
A general operator-algebraic framework for relativistic quantum theory is outlined. Some concepts of space-time structure 
are translated into algebra. Following deep results of Buchholz et al., the key role of massless modes, photons and gravitons,
and of Huygens' Principle in a relativistic quantum theory well suited to describe ``events'' and ``measurements'' 
is highlighted. In summary, a relativistic version of the ``$ETH Approach$'' to quantum mechanics is described.
\begin{center}
---
\end{center}
\textit{\,\, \,\,``Eine neue wissenschaftliche Wahrheit pflegt sich nicht in der Weise durchzusetzen, dass ihre Gegner 
\"{u}berzeugt werden und sich als belehrt erkl\"{a}ren, sondern vielmehr dadurch, dass die Gegner allm\"{a}hlich aussterben und dass die heranwachsende Generation von vornherein mit der Wahrheit vetraut ist.''} (Max Planck)

\end{abstract}

\tableofcontents

\section{Topics to be addressed}\label{Intro}
Anybody who attempts to work on the foundations -- or ``interpretation'' -- of quantum theory realizes quickly that this 
field is in a state of utmost confusion. Whether authorities in this matter or not, \textit{Richard Feynman} once said: 
``If someone tells you they understand quantum mechanics then all you’ve learned is that you’ve met a liar''; 
and \textit{Sean Carroll}, of the California Institute of Technology, in a popular article that appeared in the 
`New York Times' \cite{Carroll}, writes: ``... quantum mechanics has a reputation for being especially mysterious. 
What's surprising is that physicists seem to be O.K. with not understanding the most important theory they have. ... 
Physicists don't understand their own theory any better than a typical smartphone user understands what's going on inside the device. ... The whole thing is preposterous. Why are observations special? What counts as an ``observation'', anyway? 
When exactly does it happen? Does it need to be performed by a person? Is \textit{consciousness} somehow involved in the basic rules of reality? 
Together these questions are known as the ``measurement problem'' of quantum theory. ... '' -- Well, obviously a text like this leaves the reader in a state of bewilderment and/or anger! In the same article \textit{Caroll} also writes:
 ``You would naturally think, then, that understanding quantum mechanics would be the absolute highest 
 priority among physicists worldwide. ... Physicists, you might imagine, would stop at nothing until they truly
  understood quantum mechanics.''

Quite some time (perhaps thirty years) ago, I arrived at a conclusion similar to the one \textit{Caroll} reached 
in the last two sentences quoted above. In 2012, when I retired from my position at $ETH$ and did not have
 to make a career, anymore, I started to consider it to be one of my obligations to help removing some of the 
 confusion surrounding the foundations of quantum mechanics. I do not have any illusions about the chances 
 of success in pursuing this goal,\footnote{A recent paper of mine on the foundations of quantum mechanics triggered 
 the following comment from a ``colleague'': \textit{``Hi, again and again. How many time will you recycle your papers? 
 Cannot see (you?)  that no one is interested 
in your obscure thinking. Adding `$ETH$' will not help. You are old and essentially useless. Go fishing. Best, A.''}} 
not because it is impossible to understand quantum mechanics -- I actually think it {\bf{is possible}} -- but chiefly 
because people have so many prejudices about it.\\

Here is my credo in this endeavor:
\begin{itemize}
\item{Talking of the ``interpretation'' of a physical theory presupposes implicitly that the theory has reached its final form, but that it is not completely clear, yet, what it tells us about natural phenomena. Otherwise, we had better speak of the ``foundations'' of the theory. Quantum Mechanics has apparently not reached its final form, yet. Thus, it is not really just a matter of interpreting it, but of completing its foundations.}
\item{The only form of ``interpretion'' of a physical theory that I find legitimate and useful is to delineate approximately 
the ensemble of natural phenomena the theory is supposed to describe and to construct something resembling a
``structure-preserving map'' from a subset of mathematical symbols used in the theory that are supposed to 
represent physical quantities 
to concrete physical objects and phenomena (or events) to be described by the theory. 
Once these items are clarified the theory is supposed to provide its own ``interpretation''. (A good example is Maxwell's electrodynamics, augmented by the special theory of relativity.)}
\item{The ontology a physical theory is supposed to capture lies in \textit{sequences of events, sometimes called ``histories'',} which form the objects of series of observations extending over possibly long stretches of time and which the theory is supposed to describe.}
\item{In discussing a physical theory and mathematical challenges it raises it is useful to introduce clear concepts and basic principles to start from and then use precise and -- if necessary -- quite sophisticated mathematical tools to formulate the theory and to cope with those challenges.}
\item{To emphasize this last point very explicitly, I am against denigrating mathematical precision and ignoring or neglecting precise mathematical tools in the search for physical theories and in attempts to understand them, derive consequences from them and apply them to solve concrete problems.}
\end{itemize}

In this paper I will sketch some ideas about a formulation of {\bf{local relativistic quantum theory}} designed to describe ``events'' and, ultimately, to solve the ``measurement problem'' alluded to above. (In doing this I try to follow the credo formulated above.)
I will specifically address the following topics:
\begin{enumerate}
\item{Why is it fundamentally impossible to use a physical theory to predict the future? -- Sect. 2.}
\item{Why is quantum theory intrinsically probabilistic? -- Sect. 2.}
\item{How are ``locality'' and ``Einstein causality'' expressed in relativistic quantum theory; what is their meaning? -- Sect. 3.}
\item{What are ``events'' in quantum theory -- Sect. 4 -- and how does one describe their recording? What is meant by ``measuring a physical quantity''? -- Sect. 5.}
\item{How do \textit{states of physical systems} evolve in (space-)time, according to quantum theory? What is the probabilistic law governing their evolution? -- Sect. 4.}
\item{How does quantum theory distinguish between past and future; how does it talk about space-time? Could it be that a consistent ``Quantum Theory of Events'' must necessarily be relativistic and involve massless modes? Could it be that such a quantum theory could explain why space-time is even-dimensional and that it might incorporate gravitation as an ``emergent phenomenon''? -- Sect. 6.}
\end{enumerate}

\textit{Acknowledgements:} I am very much indebted to my collaborators on matters related to the results sketched in this paper;
among them to Philippe Blanchard, J\'{e}r\'{e}my Faupin, Martin Fraas, and especially to \textit{Baptiste Schubnel}. I also thank Detlev Buchholz, Gian Michele Graf, Klaus Hepp, Sandu Popescu, Renato Renner, but primarily Detlef D\"{u}rr and Shelly Goldstein for many helpful and enjoyable discussions and for serving as patient ``sounding boards''.

I wish to mention that various ideas related to ones elaborated on in \cite{BFS-forks, Fr2} and in this paper have been 
described in \cite{Haag, Kay}. In particular, many years ago, the late \textit{Rudolf Haag} has emphasized the importance 
of introducing a clear notion of ``events'' in quantum theory and to elucidate their role.

This paper is dedicated to the memory of \textit{Gian Carlo Ghirardi}. My approach to the foundations of quantum mechanics (dubbed \textit{``ETH Approach''}) shares some general features with $GRW$ \cite{GRW}; in particular, an important role is played by ``state collapse''. I wish to thank \textit{Detlef D\"{u}rr} for having invited me to present my ideas in this book.

\section{Why are we not able to predict the future by using our physical theories, and why is quantum theory intrinsically probabilistic?}
Imagine that the space-time of our Universe has an event horizon that hides what may happen in causally disconnected regions of space-time.
\textit{Figure 1}, below, illustrates the claim that, for fundamental reasons, 
observers are then unable to use relativistic theories to fully predict
their future; for, {\bf{never}} do they have access to complete knowledge of the initial conditions of the Universe 
that would be necessary (but not necessarily sufficient) to predict the future.\footnote{The same is true if there exist waves propagating at the speed of light along \textit{surfaces} of light-cones} This argument applies to both, 
classical \textit{and} quantum theories. But quantum theories have an additional feature that makes it impossible 
to use them to predict the future precisely: \textit{They are fundamentally probabilistic}.

\textit{Figure 1} is supposed to illustrate, furthermore, that the \textit{``Past''} consists of a ``History of Events'' or ``Facts'', while the \textit{``Future''} consists of an ensemble of ``Potentialities''. In a proper formulation of Quantum Mechanics this dichotomy should be retained! In this paper we will try to find out how to implement it in relativistic quantum theory.

\begin{center}
\includegraphics[width= 8.6cm]{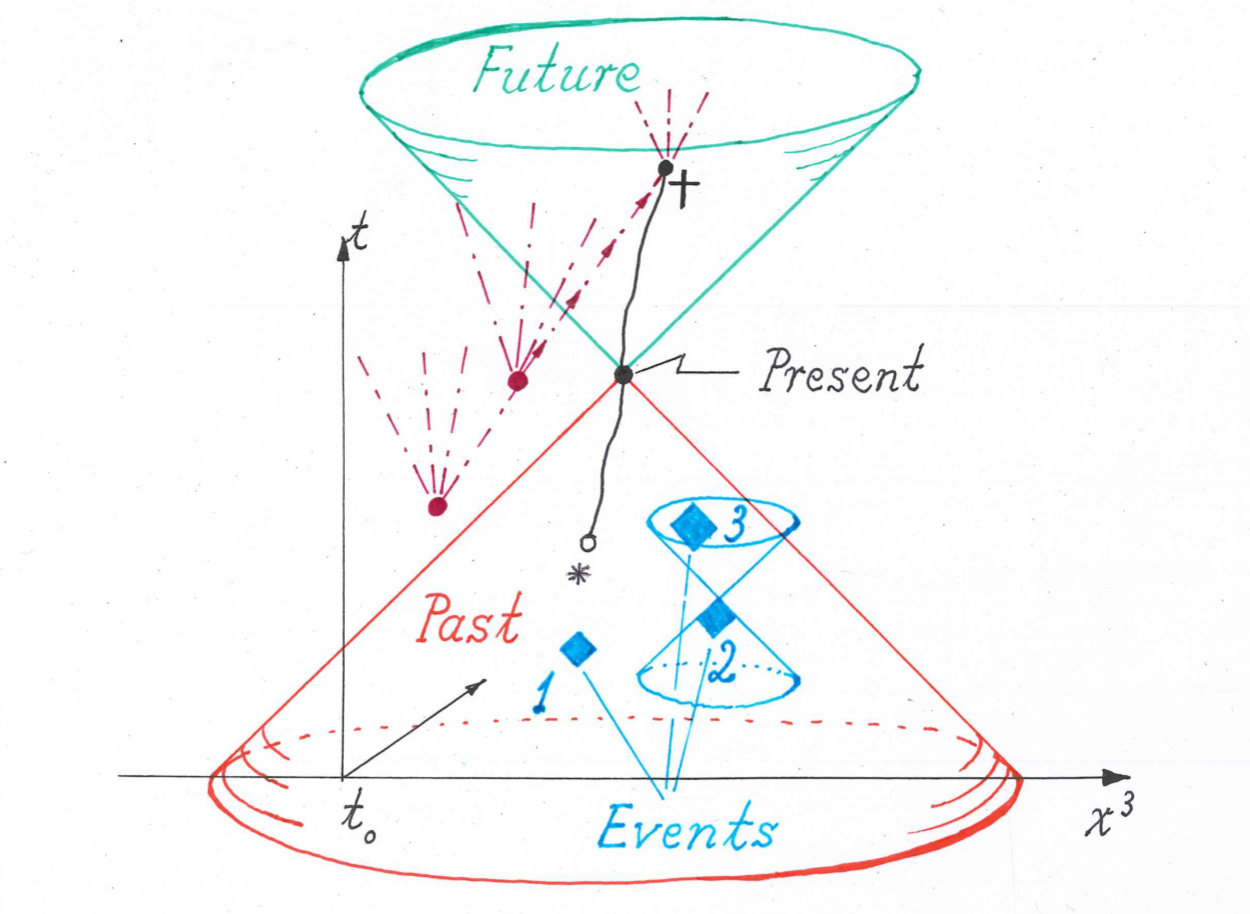}\\
\textit{Fig. 1}
\end{center}
{\small{\textit{Caption:} The ``observer'' sits at ``Present'' and is unaware of the dangers lurking from outside his past light-cone (denoted \textit{``Past''}). He might get killed at $\dagger$, a space-time point in his future light-cone (denoted \textit{``Future''}). Events are numbered in the figure; events 1 and 2 are space-like separated, event 3 is in the future of event 2.}}\\

Let $S$ be an \textit{``isolated physical system''} to be described by a model of relativistic quantum theory. -- \textit{Note:} An isolated system has the property that, over some period of time, its evolution does not depend on anything happening in its complement, i.e., in the rest of the Universe, in the sense that, during a certain period of time, the \textit{Heisenberg-picture dynamics} of physical quantities characteristic of $S$ is, \textit{for all practical purposes}, {\bf{independent}} of the degrees of freedom in the complement of $S$, (a consequence of cluster properties). It should be noted, however, that the state of $S$ can be entangled with the state of its complement!. --
 
 The concept of an \textit{isolated} physical system is important in quantum mechanics, because, \textit{only for such systems}, we know how to describe the time evolution of operators representing physical quantities in the Heisenberg picture (in terms of conjugation of those operators with the unitary propagator of the system). In order to describe the quantum dynamics of an isolated physical system $S$, we will allways start from the Heisenberg-picture dynamics of ``observables'' (i.e., of self-adjoint operators representing physical quantities) referring to $S$. The dynamics of \textit{states} of $S$ is considerably more subtle to understand and is, in a sense, at the \textit{core of our considerations} in this paper -- as it has been in the work of Ghirardi, Rimini and Weber.

In this paper we use (for simplicity) the following pedestrian formulation of the quantum mechanics of an isolated physical system $S$ in the Heisenberg picture: States of $S$ are given by density matrices, $\Omega$, acting on a separable Hilbert space, $\mathcal{H}$, of ``pure state vectors'' of $S$. 
Let $\hat{X}$ be a physical quantity of $S$, and let $X(t)=X(t)^{*}$ be the self-adjoint linear operator on $\mathcal{H}$ representing $\hat{X}$ at time $t$. Then the operators $X(t)$ and $X(t')$ representing $\hat{X}$ at two \textit{different} times $t$ and $t'$, respectively, are unitarily conjugated to one another: 
\begin{equation}\label{Heisenberg-pic}
 X(t)= U(t',t)\,X(t') U(t,t')\,,
 \end{equation}
where, for each pair of times $t, t'$, $U(t,t')$ is the propagator (from $t'$ to $t$) of the system $S$, which is a unitary operator acting on $\mathcal{H}$, and $\big\{U(t,t')\big\}_{t,t' \in \mathbb{R}}$ satisfy
 $$U(t,t')\cdot U(t',t'')=U(t, t''), \,\,\forall \text{ pairs  }t, t', \quad U(t,t) = {\bf{1}}\,, \,\forall\, t\,.$$
 
 It is often said that, in the Heisenberg picture, states of $S$ are \textit{independent} of time; and that the Heisenberg picture is equivalent to the Schr\"{o}dinger picture, where physical quantities are time-independent, but states evolve according to the propagator $U(t,t')$, solving a \textit{deterministic} Schr\"{o}dinger equation. Even if quantum mechanics were put under the auspices of the so-called ``Copenhagen interpretation'', this is, of course, nonsense, as has been amply demonstrated on many examples; (see \cite{Bell-book, FFS, Hardy}, and refs. given there)! 
 For, whenever a ``measurement'' is made, at some time $t$, say -- we will later speak, more accurately, of an {\bf{``event'' }}happening at approximately \mbox{time $t$ --} the deterministic unitary evolution of the state of $S$ in the Schr\"{o}dinger picture is {\bf{interrupted}} at this time, and the state ``jumps'', or ``collapses'' into an eigenspace of the ``observable'' that is measured -- more accurately: the state jumps into the image of an orthogonal projection representing the ``event'' that actually happens at time $t$, with jumping probabilities as given by \textit{Born's Rule}; (see also \cite{Haag, Fr2}). Expressed in the Heisenberg picture, one can say that, while operators representing physical quantities referring to an isolated physical system $S$ evolve in time according to Eq.~\eqref{Heisenberg-pic}, the \textit{state} of $S$ changes \textit{randomly} whenever an ``event'' happens; it thus exhibits a \textit{non-trivial, stochastic evolution} in time, a kind of \textit{stochastic branching process} described in \cite{FS-Prob-Th, BFS-forks, Fr2, Fr1} and in Sect.~4 of this paper. In order to avoid paradoxes \cite{Wigner, Hardy, Renner}, it is crucial to assume that the occurrence of an event (for example, the successful completion of a measurement) has an {\bf{objective}} meaning independent of the ``observer'' -- and independent of whether an ``observer'' is actually present or not.

One should think that, by now, these things are exceedingly well-known and appreciated, and hence 
I won't dwell on them any further. -- It might be added, however, that, in \textit{Bohmian mechanics}, 
randomness enters in a way that differs from the one in other formulations of quantum mechanics: 
Randomness is due, in Bohmian mechanics, to incomplete knowledge of initial conditions; see \cite{Durr-Teufel}.\footnote{The Bohmian point of view cannot be discussed any further in this paper}
 
 \section{The meaning of ``locality'' or ``Einstein causality'' in relativistic quantum theory}
In this section, I sketch remarks on ``locality'' or ``Einstein causality''. For, there appears to exist a certain amount of confusion concerning the question in which sense quantum mechanics is \textit{``non-local''} and in which sense it is perfectly \textit{``local''}. Let us consider an isolated system, $S$,  consisting of two spin-$\frac{1}{2}$ particles, $p$ and $p'$, and of equipment serving to measure components of their spins along two directions given by unit vectors $\vec{n}$ and $\vec{n}'$, respectively. 
 We imagine that, after preparation of the initial state, $\Omega$, of $S$, particle $p$ propagates into a cone, $C$, opening in the direction of the negative $x$-axis, while $p'$ propagates into a cone, $C'$, opening in the direction of the positive $x$-axis, with only tiny probabilities for sojourn outside $C$ and $C'$, respectively. Let us assume that the measurement of the spin of $p$ takes place inside a region $B \subset C$ in an interval $[t_1,t_2]$ of times, while the measurement of the spin of $p'$ takes place in a region $B' \subset C'$ within a time-interval 
 $[t'_1, t'_2]$, and let us imagine that the space-time regions $B \times [t_1,t_2]$ and $B' \times [t'_1, t'_2]$ are \textit{space-like separated}. 
 The results of the two measurements are described by two orthogonal projection operators,
 $\Pi^{p}_{\vec{n}, \sigma}, \,\sigma=\pm,$ and $\Pi^{p'}_{\vec{n}', \sigma'},\, \sigma'=\pm,$ where ``$\sigma = +$'' means that the spin of $p$ is aligned with $\vec{n}$ after the measurement has been completed, while ``$\sigma= -$'' 
 means that the spin of $p$ is anti-parallel to 
 $\vec{n}$ after its measurement, and similarly for $p'$. The operators $\Pi^{p}_{\vec{n}, \, \sigma}, \sigma= \pm,$ have the following properties:
 \begin{equation}\label{proj}
 \Pi^{p}_{\vec{n}, +}\cdot \Pi^{p}_{\vec{n}, -} =0, \,\, \Pi^{p}_{\vec{n}, +} + \Pi^{p}_{\vec{n}, -} = {\bf{1}}\,,
 \end{equation}
 and similarly for the operators $\Pi^{p'}_{\vec{n}', \sigma'}, \sigma' =\pm$. Moreover, the operators $\Pi^{p}_{\vec{n}, \sigma}$
 and $\Pi^{p'}_{\vec{n}', \sigma'}$ are localized in \textit{space-like separated} regions, $B \times [t_1,t_2]$ 
 and $B' \times [t'_1, t'_2]$, respectively, of space-time, for all choices of 
 $\sigma$ and of $\sigma'$. We would like to make an educated guess of the state used 
 by a localized observer, $\mathcal{O}$, to predict his future if $\mathcal{O}$ has the 
 property that the past light-cones of all points inside $\mathcal{O}$ contain both regions, 
 $B \times [t_1,t_2]$ and $B' \times [t'_1, t'_2]$. The answer to the question which of the two spin measurements was initiated or completed \textit{first} then obviously depends on the past ``world-tube'' of the observer $\mathcal{O}$.
 This is because 
 $B \times [t_1,t_2]$ and $B' \times [t'_1, t'_2]$ are space-like separated.
Let us suppose that, for an observer $\mathcal{O}$, the spin of $p$ was measured first, that the state of $S$ before any of these measurements were carried out was given by a density matirx $\Omega$, and that between the preparation of the state 
$\Omega$ of $S$ and further observations by $\mathcal{O}$ \textit{only} the measurements of the spins of $p$ and of $p'$ happened. According to the standard ``projection postulate'' (of the Copenhagen interpretation), the state used by $\mathcal{O}$ to predict future measurement outcomes is then given by
\begin{equation}\label{spin-meas}
\Omega_{\mathcal{O}}= [\mathcal{N}_{(\vec{n},\sigma), (\vec{n}', \sigma')}]^{-1}\, \Pi^{p'}_{\vec{n}', \sigma'}\cdot \Pi^{p}_{\vec{n}, \sigma}\, \Omega\,\, \Pi^{p}_{\vec{n}, \sigma}\cdot \Pi^{p'}_{\vec{n}', \sigma'}\,,
\end{equation}
where $\mathcal{N}_{(\vec{n},\sigma), (\vec{n}', \sigma')}:= \text{tr}\Big(\Pi^{p'}_{\vec{n}', \sigma'}\cdot \Pi^{p}_{\vec{n}, \sigma}\, \Omega\,\, \Pi^{p}_{\vec{n}, \sigma}\cdot \Pi^{p'}_{\vec{n}', \sigma'}\Big)$ is a normalization factor.
Imagine now that $\mathcal{O}'$ is an observer localized in the \textit{same} space-time region as $\mathcal{O}$, but for whom the spin of $p'$ is measured \textit{before} the spin of $p$. He then proposes to use the state $\Omega_{\mathcal{O}'}$ given by a formula arising form \eqref{spin-meas} by exchanging the order of $\Pi^{p}_{\vec{n}, \sigma}$ and $\Pi^{p'}_{\vec{n'}, \sigma'}$.
We want to impose the requirement that \textit{the predictions made by $\mathcal{O}$ and $\mathcal{O'}$ concerning future measurements} (i.e., ones localized in their common future light-cone) \textit{must be compatible}. This implies that the two states $\Omega_{\mathcal{O}}$ and $\Omega_{\mathcal{O'}}$ must agree on the algebra of all ``observables'' potentially measureable in the future of $\mathcal{O} =$ future of 
$\mathcal{O'}$. This would be guaranteed if (but does {\bf{not}} imply that)
\begin{equation}\label{locality}
\Pi^{p'}_{\vec{n}', \sigma'}\cdot \Pi^{p}_{\vec{n}, \sigma}=\Pi^{p}_{\vec{n}, \sigma}\cdot \Pi^{p'}_{\vec{n}', \sigma'}\,,
\end{equation}
for arbitrary choices of $(\vec{n}, \sigma)$ and $(\vec{n}' , \sigma')$, assuming, as stated above, that the localization regions $B \times [t_1,t_2]$ and $B' \times [t'_1,t'_2]$ are space-like separated. Equation \eqref{locality} is what is called \textit{``locality''} or \textit{``Einstein causality''} in relativistic quantum field theory. 
This is a sufficient (but not necessary) condition to eliminate ambiguities in the predictions of possible future measurement outcomes made by different observers that are due to the impossibility of unambiguously ordering measurements according to the times at which they are initiated (or completed). But Eq. \eqref{locality} does {\bf{not}} imply that quantum mechanics is ``local'' in the following sense:
Consider the state 
$$\Omega_{(\vec{n}, \sigma)}:= [\mathcal{N}_{(\vec{n},\sigma)}]^{-1} \Pi^{p}_{\vec{n}, \sigma}\, \Omega\,\, \Pi^{p}_{\vec{n}, \sigma},$$
where $\mathcal{N}_{(\vec{n},\sigma)}$ is a normalization factor chosen such that tr$(\Omega_{(\vec{n}, \sigma)})=1$. Let $A$ be an ``observable'' localized in a space-time region space-like separated from 
$B \times [t_1,t_2]$; (for example $A=\Pi^{p'}_{\vec{n}', \sigma'}$). One might expect that 
$$\text{tr}\big(\Omega \,A\big) = \text{tr}\big(\Omega_{(\vec{n}, \sigma)} \, A\big)\,,$$
for any operator $A$ with these properties.
But, of course, this equality does {\bf{not}} hold! This fact is what people call the ``non-locality'' of quantum theory. 
In quantum field theory, this kind of ``non-locality'' is neatly reflected in the Reeh-Schlieder theorem \cite{Jost}. It results from entanglement.

One major purpose of this paper is to render the ``projection postulate'' (or ``collapse postulate'' -- see Eq.~\eqref{spin-meas}) more precise, to explain its origin and to find out \textit{under what conditions it is applicable}. In contrast to the ideas described in \cite{GRW}, we will not invoke any mechanism extraneous to quantum mechanics that produces ``state collapse''.

\section{Relativistic quantum theory, and the notion of ``events''}

In this section we propose an algebraic definition of local relativistic quantum theory and then introduce a precise notion of 
``events''. We require some rudimentary knowledge of the theory of operator algebras. In particular, the reader might profit from knowing what a $C^{*}$- and what a von Neumann algebra is and what, for example, the Gel'fand-Naimark-Segal ($GNS$) construction is. What will be used from the theory of operator algebras, in this paper, can be learned in a few hours! A useful reference may be \cite{Brat-Rob}.

For the time being, we will consider space-time, $\mathcal{M}$, to be given; but we do not equip $\mathcal{M}$ with a Lorentzian metric. Later, we will try to clarify how properties of algebras of operators representing localized potentialities 
equip $\mathcal{M}$ with a causal structure. But to start with, we assume $\mathcal{M}$ to be given by Minkowski space, 
$\mathbb{M}^{d}$, with $d=4$. 

In relativistic quantum theory, all operators representing physical quantities characteristic of an isolated physical system $S$ can be localized in some space-time regions.
Given a region $\mathcal{O}\subset \mathcal{M}$, we denote by $\mathcal{A}(\mathcal{O})$ the algebra generated by
all bounded operators localized in $\mathcal{O}$ that represent physical quantities. The family $\big\{ \mathcal{A}(\mathcal{O})\big\}_{\mathcal{O} \subset \mathcal{M}}$ is called a ``net of local algebras''. For an introduction to these concepts and to algebraic quantum field theory the reader is advised to consult \cite{Haag1}. In the following considerations, the regions 
$\mathcal{O}$ are usually taken to be forward or backward light-cones with apex in an arbitrary space-time point $P\in \mathcal{M}$. \\

\underline{A general formulation of local relativistic quantum theory}:\\
 
 We consider an isolated physical system $S$ to be described with the help of a model of local relativistic quantum theory.\\
 
\underline{Definition 1}: By $\mathcal{F}_{P}$ we denote the $^{*}$algebra generated by all operators representing physical quantities referring to $S$ (such as potential events) localized in the ``future'' of the space-time point $P$, while $\mathcal{P}_{P}$ denotes the algebra generated by all operators representing physical quantities localized in the ``past'' of $P$. \hspace{9.1cm}$\Box$\\

We assume that all the algebras $\mathcal{F}_{P}$ are contained in a $C^{*}$-algebra $\mathcal{E}$, and
\begin{equation}\label{quasi-local alg}
\mathcal{E} = \overline{\bigvee_{P \in \mathcal{M}} \mathcal{F}_{P}}\,,
\end{equation}
where the closure on the right side is taken in the operator norm of $\mathcal{E}$. 
We assume that all these algebras are represented on a common separable Hilbert space $\mathcal{H}$ and that all \textit{``states of physical interest''} of $S$ can be identified with density matrices (non-negative trace-class operators normalized to have trace $=1$) acting on $\mathcal{H}$.\footnote{It is sometimes advantageous to formulate this assumption 
in a more abstract, algebraic way involving, among other ingredients, the $GNS$-construction; see, e.g., \cite{Haag1}.} In our notation, we will not distinguish between an abstract element of the algebra $\mathcal{E}$ and the linear operator on 
$\mathcal{H}$ representing it.\\

\underline{Definition 2}: We define $\mathcal{E}_{P}$ to be the von Neumann algebra obtained by closure of the algebra 
$\mathcal{F}_{P}$ in the weak operator topology of the algebra, $B(\mathcal{H})$, of all bounded operators on $\mathcal{H}$.
\hspace{6.4cm} $\Box$\\

If $S$ is a physical system in a state of \textit{finite} energy describing only excitations of strictly positive rest mass then 
\begin{equation}\label{No-DP}
\mathcal{E}_{P} \simeq B(\mathcal{H})\,, \text{  for any point  }\, P\in \mathcal{M}\,.
\end{equation}
It is expected that this equality always holds in a space-time of \textit{odd} dimension, \textit{even} if massless particles are present. This is because Huygens' Principle does not hold in space-times of odd dimension. (It also does not hold in certain even-dimensional space-times with non-vanishing curvature. But that's another story, which, for reasons that 
I will not explain in any detail, is not expected to invalidate the following considerations.) The property 
expressed in Eq. \eqref{No-DP} is one most people sub-consciously consider to be always valid. 
But this is actually \textit{not} the case! (If it were we would probably be unable to introduce a 
reasonable notion of ``events'' in quantum theory, and we would never solve the ``measurement problem''.)

If there exist massless particles, in particular photons and/or gravitons and Dark-Energy modes, and if Huygens' Principle holds in an appropriate sense ($\mathcal{M}$ even-dimensional, specifically 
$\mathcal{M}=\mathbb{M}^{4}$),\footnote{or in the presence of blackholes in space-time} the algebra $\mathcal{E}_{P}$ 
tends to have an 
infinite-dimensional commutant, $\mathcal{E}_{P}'$. (The commutant, $\textfrak{M}'$, of an algebra $\textfrak{M}$ contained in $B(\mathcal{H})$ is the algebra of all bounded operators on $\mathcal{H}$ commuting with \textit{all} 
operators in $\textfrak{M}$.) 
More specifically, within an algebraic framework of local relativisitic quantum field theory over four-dimensional 
Minkowski space-time, \textit{Detlev Buchholz} has shown \cite{Buch1} that, in the presence of massless particles,
$\mathcal{E}'_{P_t} \cap \mathcal{E}_{P_{t_0}} $ is an \textit{infinite-dimensional, non-commutative} algebra, whenever $P_{t_0}$ is a space-time point in the \textit{past} of the space-time point $P_t$, as indicated in Figure 2. 

In his proof, \textit{Huygens' Principle} is exploited in the form that asymptotic out-fields creating on-shell massless particles 
escaping to infinity do not propagate into the \textit{interior} of forward light-cones contained in the future 
of the space-time region (denoted by $\mathcal{O}$ in Figure 2) where they are localized, but propagate 
along the \textit{surface} of forward light-cones with apices in $\mathcal{O}$. Such asymptotic out-fields are then shown to commute 
with all operators in the algebra $\mathcal{E}_{P_t}$.

\begin{center}
\includegraphics[width=7cm]{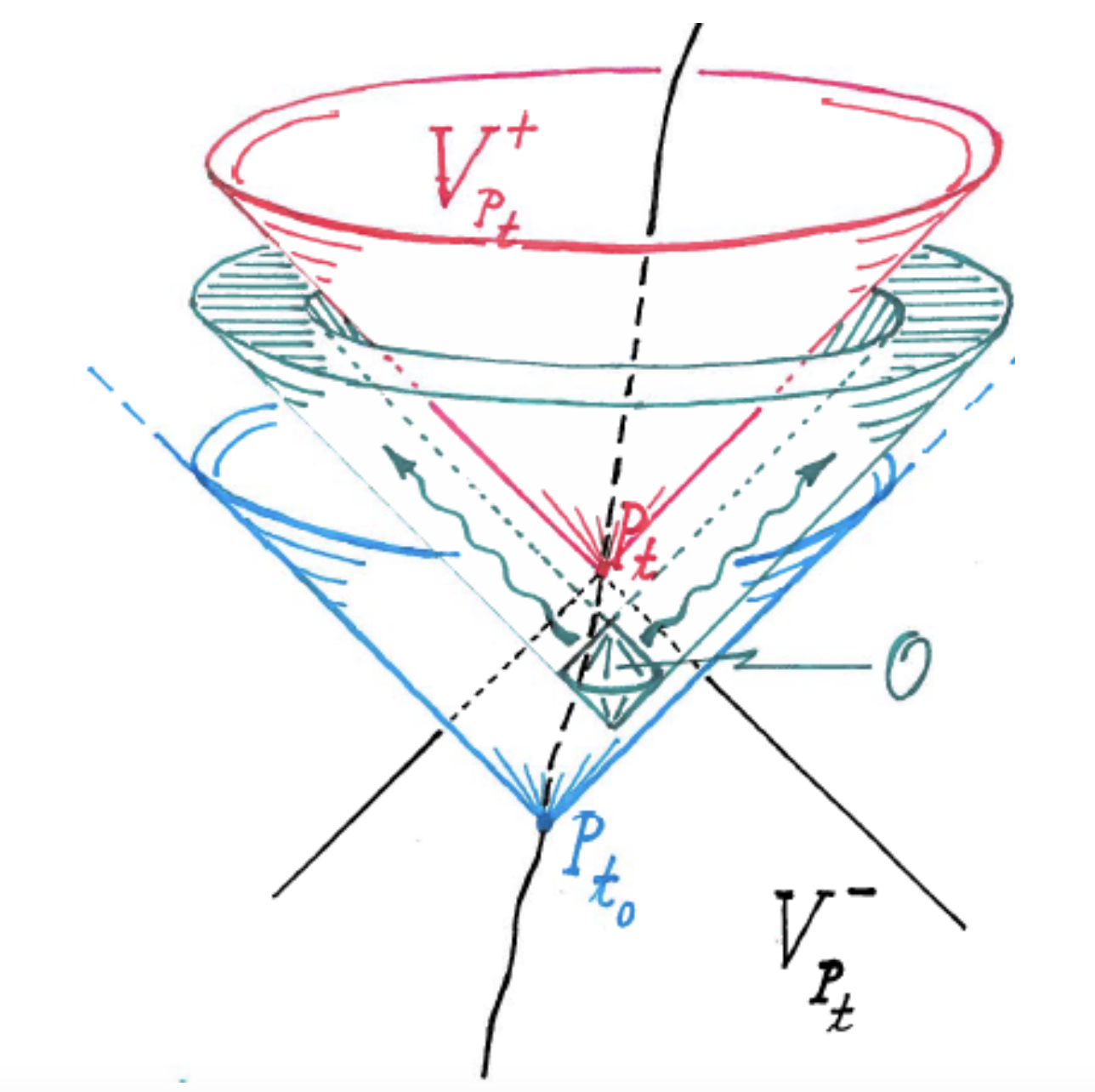}\\
\textit{Fig. 2}
\end{center}
{\small{\textit{Caption:} The black line is the world-line of an ``observer'' who, at time $t$, is localized near $P_{t}$. Operators representing physical quantities potentially observable by the ``observer'' in the future of $P_t$ are localized 
inside the forward light-cone $V^{+}_{P_t}$. They generate the algebra $\mathcal{E}_{P_t}$. Asymptotic out-field operators
describing the emission of (on-shell) photons or gravitons in the region $\mathcal{O}$ propagate along the light-cones contained in $V^{+}_{P_{t_0}}$ but \textit{not} contained in $V^{+}_{P_t}$.}}\\

One expects that, if space-time is even-dimensional and in the presence of massless particles, the algebras $\mathcal{E}_{P}$ have the property that all non-zero orthogonal projections belonging to $\mathcal{E}_{P}$ have an infinite-dimensional range. 
This implies that there do {\bf{not}} exist any normal \textit{pure} states on these algebras. Furthermore, they are expected to be isomorphic to a certain ``universal'' von Neumann algebra, $\textfrak{N}$,\footnote{\, $\textfrak{N}$ is expected to be a von Neumann algebra of type $III_{1}$} i.e., 
$\mathcal{E}_{P} \simeq \textfrak{N}, \, \forall\, P\in \mathcal{M}$.

We now use these insights to extract a {\bf{general algebraic formulation}} of \textit{local relativistic quantum theory} compatible with the appearence of ``events'' and promising a solution of the ``measurement problem''.
We assume that space-time $\mathcal{M}$ is a topological space with the property that, with every point $P \in \mathcal{M}$, one can associate a von Neumann algebras, $\mathcal{E}_{P}$, the ``algebra of potential events that might possibly happen in the future of $P$\,'', with the property that $\mathcal{E}_{P}$ is contained in a $C^{*}$-algebra $\mathcal{E}$, for all $P\in \mathcal{M}$. 

The family of algebras $\big\{\mathcal{E}_{P}\big\}_{P\in \mathcal{M}}$ equips space-time $\mathcal{M}$ with the following \textit{causal structure}:

\underline{Definition 3}:
A space-time point $P'$ is in the future of a space-time point $P$, written as $P'\succ P$, (or, equivalently, $P$ is in the past of $P'$, written as $P\prec P'$\,) iff
\begin{equation}\label{PDP}
\boxed{
\,\mathcal{E}_{P'} \subsetneqq \mathcal{E}_{P}\,, \quad \mathcal{E}'_{P'} \cap \mathcal{E}_{P} \,\,\text{is an } \infty-\text{dim. non-commutative algebra}\,
}
\end{equation}
\hspace{11.8cm} $\Box$\\
Equation \eqref{PDP} expresses what I call the
\begin{center}
\textit{``Principle of Diminishing Potentialities'' (PDP)}
\end{center}
This principle is a {\bf{theorem}} in an axiomatic formulation of quantum electrodynamics over four-dimensional Minkowski space proposed by
D. Buchholz and the late J. Roberts \cite{BR}. 

Henceforth, the \textit{Principle of Diminishing Potentialities 
will always be {\bf{assumed}} to hold}; and, within our formulation of relativistic quantum theories, (a model of ) an \textit{isolated physical system} $S$ is defined by specifying the following data: 
\begin{equation}\label{systems}
S=\big\{\mathcal{M}, \mathcal{E}, \mathcal{H}, \big\{\mathcal{E}_{P}\big\}_{P\in \mathcal{M}}\, \text{ satisfying }\,PDP\big\}\,,
\end{equation}
where $\mathcal{M}$ is a model of space-time, $\mathcal{E}$ is a $C^{*}$-algebra represented on a Hilbert space 
$\mathcal{H}$, and $\big\{ \mathcal{E}_{P} \big\}_{P \in \mathcal{M}}$ is a family of von Neumann algebras satisfying the ``Principle of Diminishing Potentialities'' introduced in Eq. \eqref{PDP}. \\

\underline{Definition 4}:
If a space-time point $P'$ is neither in the future of a space-time point $P$ nor in the past of $P$ we say that $P$ and $P'$ are \textit{space-like separated}, written as
$P$\, {\Large$\bigtimes$} \,$P'$. \hspace{6.4cm} $\Box$\\

Let $\Sigma$ be a space-like subset of $\mathcal{M}$. If $\mathcal{M}= \mathbb{M}^{4}$ we imagine that $\Sigma$ is a subset of a space-like hypersurface of co-dimension 1 in $\mathcal{M}$. Since all the algebras 
$\mathcal{E}_{P}, p \in \mathcal{M},$ are assumed to be contained in the $C^{*}$-algebra 
$\mathcal{E}$, the following definition is meaningful:
\begin{equation}\label{algebra}
\mathcal{E}_{\Sigma}:= \overline{\bigvee_{P\in \Sigma} \mathcal{E}_{P}}\,,
\end{equation}
where the closure is taken in the weak topology of $B(\mathcal{H})$.
A state,  $\omega_{\Sigma}$, on the algebra $\mathcal{E}_{\Sigma}$ is a normalized, positive linear functional on $\mathcal{E}_{\Sigma}$.
    
\textit{Remark:} At this point we should comment on the question of what the operational meaning of a ``state'' of an isolated system $S$ is, and how one can \textit{prepare} $S$ in a specific state. Obviously these are important questions, which, however, cannot be discussed here; but see \cite{FS-state-prep}.\\

\underline{Definition 5}: Let $\textfrak{M}$ be a von Neumann algebra, and let $\omega$ be a normal state on $\textfrak{M}$. For an operator $X\in \textfrak{M}$, we define $ad_{X}(\omega)$ to be the linear functional on $\textfrak{M}$ defined by
$$ad_{X}(\omega)(Y):= \omega([Y,X]), \quad \forall \,Y \in \textfrak{M}.$$
We define the \textit{centralizer}, $\mathcal{C}_{\omega}(\mathfrak{M})$, of the state $\omega$ by
\begin{equation}\label{centralizer}
\mathcal{C}_{\omega}(\mathfrak{M}):=\big\{X\,\vert X \in \mathfrak{M}, ad_{X}(\omega)=0\big\}.
\end{equation}
It is easy to verify that $\mathcal{C}_{\omega}(\mathfrak{M})$ is a (von Neumann) subalgebra of $\textfrak{M}$, and that $\omega$ is a normalized trace on $\mathcal{C}_{\omega}(\mathfrak{M})$. (This property implies that centralizers are completely classified!)\\
Given an algebra $\textfrak{N}$, the \textit{center}, $\mathcal{Z}(\textfrak{N}),$ is the abelian subalgebra of $\textfrak{N}$ consisting of all operators in $\textfrak{N}$ commuting with all other operators in $\textfrak{N}$. We set
\begin{equation}\label{center}
\mathcal{Z}_{\omega}(\textfrak{M}):= \mathcal{Z}(\mathcal{C}_{\omega}(\mathfrak{M}))
\end{equation}
\hspace{11.8cm} $\Box$

Motivation underlying the following notions and definitions is provided in \cite{BFS-forks, Fr2, Fr1}. \\

\underline{Definition 6}: Given a point $P \in \mathcal{M}$, a \textit{potential event} in the future of $P$ is a family,
$\big\{ \pi_{\xi} \vert\, \xi \in \textfrak{X}\big\}$, ($\textfrak{X}$ a countable set of indices\footnote{Here it is assumed that potential events can be identified with the spectral projections of self-adjoint operators with \textit{discrete} spectrum 
($\simeq \textfrak{X}$); more generally, one could identify potential events with spectral projections of families (abelian algebras) of commuting self-adjoint operators that may have continuous spectrum}), of orthogonal projections belonging to 
$\mathcal{E}_{P}$ with the properties
\begin{equation}\label{pot-event}
\pi_{\xi}\cdot \pi_{\eta}= \delta_{\xi \eta} \pi_{\xi}, \,\,\forall \,\, \xi, \eta \in \textfrak{X}, \quad \sum_{\xi \in \textfrak{X}} \pi_{\xi} = {\bf{1}}\,.
\end{equation}
It is expected that events usually have a \textit{finite duration}. This would imply that operators 
$\big\{ \pi_{\xi} \vert \xi \in \textfrak{X}\big\}$ representing a potential event in the future of the point $P$ would be localized in a \textit{compact} region of space-time contained in the future of $P$ (the future light-cone with apex in $P$). \hspace{2.5cm}$\Box$\\

\underline{Definition 7}: Given a state $\omega_{P}$ on the algebra $\mathcal{E}_{P}$, we say that an \textit{event happens in the future of the space-time point $P$} iff the algebra 
$$\mathcal{Z}_{\omega_P}:= \mathcal{Z}\big(\mathcal{C}_{\omega_P}(\mathcal{E}_{P})\big)$$
is generated by the projections $\big\{ \pi_{\xi} \vert\, \xi \in \textfrak{X}\big\} \subset \mathcal{Z}_{\omega_P} \subset \mathcal{E}_{P}$ of a potential event in the future of $P$ with the properties that the cardinality of $\textfrak{X}$ is at least 2 and that there exist projections $\pi_{\xi_1}, ... , \pi_{\xi_n}, n\geq 2,$ such that
\begin{equation}\label{Born1}
\omega(\pi_{\xi_j}) > 0, \qquad \forall j=1, ... , n, \,\, n\geq 2\,.
\end{equation}
(The quantity $\omega(\pi_{\xi})$ will turn out to be the \textit{Born probability} for $\pi_{\xi}$ to occur in the future of $P$.)
\hspace{8.5cm}$\Box$\\

Let $\omega_P$ be the state of $S$ on the algebra $\mathcal{E}_{P}$. It is easy to see that if an event described by the family 
$\big\{ \pi_{\xi} \vert\, \xi \in \textfrak{X}\big\}\subset \mathcal{Z}_{\omega_P}$ of projections 
happens in the future of the point $P$ then
\begin{equation}\label{mixture}
\omega_{P}(X)= \sum_{\xi \in \textfrak{X}} \omega\big( \pi_{\xi} \, X \, \pi_{\xi} \big), \quad \forall X \in \mathcal{E}_{P}\,,
\end{equation}
 i.e., the state $\omega_P$ on the algebra $\mathcal{E}_{P}$ is a \textit{mixture} of the states 
 \begin{equation}\label{collapse}
 \omega_{P,\xi}:=\big[\omega_{P}(\pi_{\xi})\big]^{-1} \omega\big(\pi_{\xi}(\cdot) \pi_{\xi}\big)
 \end{equation}
 labelled by the points $\xi \in \textfrak{X}$.\\

The following is a crucial axiom.\\

\underline{\bf{Axiom 1}} (``State-collapse'' postulate): If an event happens in the future of a point $P\in \mathcal{M}$, in the sense of Definition 7, then the state
to be used to make predictions of further events possibly happening in the future of $P$ is given by $\omega_{P,\xi_{*}}$, 
for some $\xi_{*} \in \textfrak{X}$ with $\omega_{P}(\pi_{\xi_{*}})>0$, where $\omega_{P,\xi_{*}}, \xi_{*}\in \textfrak{X},$ is defined in Eq. \eqref{collapse}.

The probability that $\omega_{P, \xi_{*}}$ is selected among the states $\big\{\omega_{P, \xi}\vert\, \xi \in \textfrak{X} \big\}$ is given by \textit{Born's Rule}, namely it is given by $\omega_{P}(\pi_{\xi_{*}})$. The projection $\pi_{\xi_{*}}$ is called the \textit{``actual event''} happening in the future of $P$.\hspace{2.6cm} $\Box$\\

Next, we consider two points, $P$ and $P'$, in a subset $\Sigma$ of $\mathcal{M}$, with $P\bigtimes P'$, (i.e., $P$ and $P'$  are space-like separated), We assume that the state 
$\omega_{\Sigma}$ defined in Eq. \eqref{algebra} is given, so that the states 
$\omega_{P}=\omega_{\Sigma}\vert_{\mathcal{E}_P}$ and $\omega_{P'}=\omega_{\Sigma}\vert_{\mathcal{E}_P'}$
are known, too. We suppose that, given $\omega_{\Sigma}$, events happen in the future of $P$ and of $P'$. Let 
$\mathcal{Z}_{\omega_P}$ denote the center of the centralizer of the state $\omega_P$ on the algebra $\mathcal{E}_P$, which describes the event $\big\{\pi^{P}_{\xi} \vert \xi \in \textfrak{X}^{P} \big\}$ happening in the future of $P$, and let 
$\mathcal{Z}_{\omega_{P'}}$ be the algebra describing the event happening in the future of the point $P'$. We require the following axiom.\\

\underline{\bf{Axiom 2}} (Events in the future of space-like separated points commute): Let $P\, \bigtimes \, P'$. Then all operators in $\mathcal{Z}_{\omega_P}$ commute with all operators in $\mathcal{Z}_{\omega_{P'}}$. In particular,
$$\hspace{2.2cm} \big[\pi^{P}_{\xi}, \pi^{P'}_{\eta}\big] =0, \,\,\,\forall\, \xi \in \textfrak{X}^{P}  \text{ and all }\, \eta \in \textfrak{X}^{P'}. \hspace{2.3cm} \Box $$\\
\indent This axiom may be one reflection of what people sometimes interpret as the fundamental {\bf{non-locality}} of quantum theory: Projection operators representing events in the future of two \textit{space-like separated} points $P$ and $P'$ in space-time are {\bf{constrained}} to commute with each other! Actually, this implies what in quantum field theory is understood to express {\bf{locality}} or Einstein causality.\\

Next, we assume that some slice, $\textfrak{F}$, in space-time $\mathcal{M}$ is foliated by space-like hypersurfaces, 
$\Sigma_{\tau}$:\,
$\textfrak{F}:=\big\{\Sigma_{\tau}\vert \tau \in [0,1]\big\}$, where $\tau$ is a time coordinate in the space-time region filled by 
$\textfrak{F}$. Let $P$ be an arbitrary space-time point in the leaf $\Sigma_{1}$, and let the ``recent past'' of $P$, $V^{-}_{P}(\textfrak{F})$, consist of all points in $\bigcup_{\tau < 1} \Sigma_{\tau}$ that are in the \textit{past} of $P$, in the sense specified in Definition 3, above. The task we propose to tackle is the following one: We suppose that we know the state 
$\omega_{\Sigma_{0}}$ on the algebra $\mathcal{E}_{\Sigma_0}$, (see Eq. \eqref{algebra}). 
Assuming that Axioms 1 and 2 hold, we propose to determine the state $\omega_{P}$ on $\mathcal{E}_{P}$, for the given point 
$P\in \Sigma_{1}$. Let 
$\big\{P_\iota \vert \iota \in \textfrak{I}(\textfrak{F})\big\}$ denote the subset of points in $V^{-}_{P}(\textfrak{F})$ in whose future events happen (see Definition 7), and let 
$$\big\{ \pi_{\xi_\iota}^{P_\iota} \vert \iota \in \textfrak{I}(\textfrak{F})\big\} \subset \mathcal{E}_{\Sigma_0}$$ 
be the \textit{actual events} (see Axiom 1) that happen in the future of the points $P_\iota\,, \iota \in \textfrak{I}(\textfrak{F})$; 
(here $\textfrak{I}( \textfrak{F})$ is a set of indices labelling the points in $V^{-}_{P}(\textfrak{F})$ in whose future events happen; it is here assumed to be countable). 
We define a so-called \textit{``History Operator''}
\begin{equation}\label{History}
H\big(V^{-}_{P}(\textfrak{F})\big):= \vec{\Pi}_{\iota \in \textfrak{I}(\textfrak{F})}\, \pi_{\xi_\iota}^{P_\iota}\,,
\end{equation}
where the ordering in the product $\Vec{\Pi}$ is such that a factor $\pi_{\xi_\kappa}^{P_\kappa}$ corresponding to a point $P_{\kappa}$ stands to the right of a factor $\pi_{\xi_\iota}^{P_\iota}$ corresponding to a point $P_{\iota}$ iff $P_{\kappa} \prec P_{\iota}$, (i.e., if $P_{\kappa}$ is in the past of $P_{\iota}$). But if $P_{\iota} \bigtimes P_{\kappa}$, i.e., if $P_{\iota}$ and $P_{\kappa}$ are space-like separated the order of the two factors is \textit{irrelevant} -- \textit{thanks to} Axiom 2!

The state on the algebra $\mathcal{E}_{P}$ relevant to make predictions about events happening in the future of $P$, in the sense of Definition 7, is then given by
\begin{equation}\label{state-propagation}
\omega_{P}(X) \equiv \omega_{P}^{\textfrak{F}}\big(X \big) = \big[\mathcal{N}_{P}^{\,\textfrak{F}}\, \big]^{-1} 
\omega_{\Sigma_0}\big(H(V^{-}_{P}(\textfrak{F}))^{*} \,X\, H(V^{-}_{P}(\textfrak{F}))\big)\,, \,X\in \mathcal{E}_{P}\,,
\end{equation}
where the normalization factor $\mathcal{N}_{P}^{\,\textfrak{F}}$ is given by
\begin{equation}\label{hist-prob}
\mathcal{N}_{P}^{\,\textfrak{F}}= \omega_{\Sigma_0}\big(H(V^{-}_{P}(\textfrak{F}))^{*} \cdot H(V^{-}_{P}(\textfrak{F}))\big)\,.
\end{equation}

We recall that, according to Definition 7, an event happens in the future of a point $P\in \Sigma_{1}$ iff the center,
$\mathcal{Z}_{\omega_{P}}$, of the centralizer of the state $\omega_{P}$ on the 
algebra $\mathcal{E}_{P}$, defined in \eqref{state-propagation}, contains at least two disjoint orthogonal projections of strictly positive probability, as given by \textit{Born's Rule}; (see Axiom 1).

The quantities $\mathcal{N}_{P}^{\,\textfrak{F}}$ can be used to equip the \textit{tree-like} space (the so-called ``non-commutative spectrum'' of $S$) of all possible histories of events in the future of points belonging to the foliation $\textfrak{F}$ with a \textit{probability measure}; see \cite{Fr2}.

The ideas and results discussed here are illustrated in Figure 3, below.

\begin{center}
\includegraphics[width=8cm]{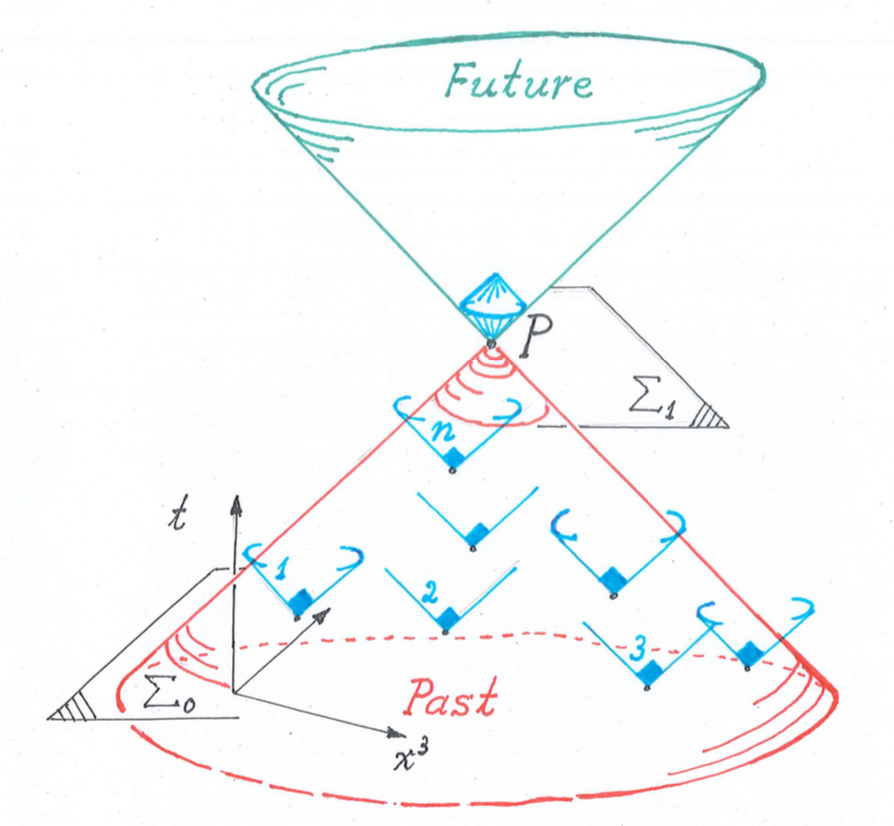}\\
\textit{Fig. 3}
\end{center}
{\small{\textit{Caption:} It is tacitly assumed here that all events that happened in the past of the point $P$ have a strictly finite duration. They are marked by small ``diamonds'' and are numbered from $1$ to $n$. Notice that $1 \bigtimes 2$ and $2 \prec n$.}}\\

To conclude this discussion, in the approach to relativistic quantum theory presented in this paper (called ``$ETH Approach$''), the {\bf{evolution}} (along the foliation 
$\textfrak{F}$) of the {\bf{state}} of an isolated physical system $S$, given the initial state 
$\omega_{\Sigma_0}$ on the algebra $\mathcal{E}_{\Sigma_0}$ defined in Eq.\eqref{algebra},\footnote{and assuming the axiom of choice} can be viewed as a \textit{generalized stochastic branching process},
whose state space is what I have called the ``non-commutative spectrum'' of the system $S$, (see \cite{Fr2}, and Eq.~\eqref{NC-spectrum}, Sect.~6, for a definition), and with \textit{branching rules} derived from Definition 7, Axioms 1 and 2 and Eqs.~\eqref{History} - \eqref{hist-prob}.\footnote{This picture has reminded my former student \textit{P.-F. Rodriguez} of the sentence from the short story \textit{``The Garden of Forking Paths''}, by \textit{Jorge Luis Borges}, that I have appended to the abstract of this paper}

Mathematical details can be made precise if space-time is discretized. Additional information can be found in \cite{Fr2, Schubnel-thesis, Les-Diablerets}.

\section{Monitoring events by measuring physical quantities}
Let $S=\big\{\mathcal{M}, \mathcal{E}, \mathcal{H}, \big\{\mathcal{E}_{P}\big\}_{P\in \mathcal{M}}\, \text{ satisfying }\,PDP\big\}$ be the data defining an isolated physical system, with the properties specified in Sect.~4, Eq.~\eqref{systems}, and assumed to satisfy Axioms 1 and 2. In Sect. 4, we have introduced a precise notion of {\bf{``events''}} featured by $S$. In this section, we propose to explain how events can be recorded/monitored by measuring physical quantities referring to $S$.

For the purposes of the present exposition it is convenient to define a \textit{``physical quantity''} to be an abstract self-adjoint linear operator $\hat{X}$ with the property that, for every point $P\in \mathcal{M}$, there exists a concrete self-adjoint linear operator $X(P) \in \mathcal{E}_P$ acting on the Hilbert space $\mathcal{H}$ of $S$ and representing the quantity $\hat{X}$; (see \cite{Fr2} for a somewhat more general and abstract notion of physical quantities). 

\textit{Remark:} If space-time $\mathcal{M}$ is given by Minkowski space $\mathbb{M}^{4}$ the operator $X(P)$ is conjugated to the operator $X(P')$ by a unitary operator on the Hilbert space $\mathcal{H}$ representing the space-time translation from $P$ to $P'$. But on general space-times a simple relation between $X(P)$ and $X(P')$ may not exist.  

We define
\begin{equation}\label{phys-quantities}
\mathcal{O}_{S}:= \Big\{ \hat{X}_{\iota}= \hat{X}_{\iota}^{*}\, \vert \,\iota \in \textfrak{I}(S) \Big\}
\end{equation}
to be a list of all physical quantities available, at present, to characterize properties of $S$ for which there exists a prescription of how they can be measured.\footnote{For simplicity, we assume that all operators in $\mathcal{O}_{S}$ have discrete spectrum} The list $\mathcal{O}_S$ is not intrinsic to the theoretical description of the system $S$; rather it specifies those physical quantities referring to $S$ that, during a given era, can be expected to be measurable in real experiments. In quantum theory, this list is \textit{not} an algebra (unless all operators belonging to $\mathcal{O}_S$ commute with one another), and it is usually not even a real linear vector space. The question to be addressed in the following is what we mean by saying that some quantity $\hat{X}\in \mathcal{O}_S$ is measured in the future of a space-time point $P$, and how such a measurement can be used to record an event that happens in the future of $P$.

Suppose that, for some point $P\in \mathcal{M}$, the \textit{center} $\mathcal{Z}_{\omega_P}$ (of the centralizer 
$\mathcal{C}_{\omega_P}(\mathcal{E}_P) \subset \mathcal{E}_P$ of the state $\omega_P$ on the algebra $\mathcal{E}_P$) is non-trivial and is generated by a family $\big\{ \pi_{\xi} \vert\, \xi \in \textfrak{X} \big\}$ of disjoint orthogonal projections describing an event happening in the future of $P$. Let $\varepsilon$ be a positive number; (it will turn out to be a measure of the ``resolution'' of the recording of this event in a measurement of a physical quantity $\hat{X} \in \mathcal{O}_S$). We let 
$\big\{\pi_1, \dots, \pi_N\big\}$ be a finite number of disjoint orthogonal projections contained in $\mathcal{Z}_{\omega_P}$ with the property that 
\begin{equation}\label{lower-bound}
\omega(\pi_j)\geq \varepsilon,\,\, \forall\, j=1, ..., N,\qquad \omega({\bf{1}}-\sum_{i=1}^{N} \pi_{i})< \varepsilon\,.
\end{equation}
The projections $\big\{\pi_1, \dots, \pi_N\big\}$ form the basis of an $N$-dimensional vector space, 
$\mathcal{V}_{\omega_P}^{(\varepsilon)}$, equipped with a (positive-definite) scalar product, 
$\langle \cdot,\cdot \rangle$, given by
\begin{equation}\label{scalar-prod}
\langle \pi_i, \pi_j \rangle := \omega(\pi_{i}\cdot \pi_{j}) = \omega(\pi_i) \,\delta_{ij} \geq  \varepsilon\,\delta_{ij}, \,\, \text{ for }\, i,j = 1,..., N\,.
\end{equation}
Every vector $Z\in \mathcal{V}_{\omega_P}^{(\varepsilon)}$ can be represented as a linear combination,
\begin{equation}\label{vectors}
Z=\sum_{j=1}^{N} z_{j} \pi_{j} \in \mathcal{Z}_{\omega_P} \,, \text { for complex numbers }\, z_1,..., z_N\,.
\end{equation}
We can thus identify $\mathcal{V}_{\omega_P}^{(\varepsilon)}$ with an $N$-dimensional subspace, 
actually an $N$-dimensional \textit{subalgebra} of $\mathcal{Z}_{\omega_P}$. 

Let  $\mathcal{H}_{\omega_P}$ be the Hilbert space and $\Omega_{P}$ the cyclic vector in $\mathcal{H}_{\omega_P}$
obtained by applying the Gel'fand-Naimark-Segal construction to the pair $\big( \mathcal{E}_P, \omega_P\big)$; (see. e.g., \cite{Brat-Rob}). 
There is a bijection between the vector space $\mathcal{V}_{\omega_P}^{(\varepsilon)}$ and the subspace $\mathcal{W}_{\omega_P}^{(\varepsilon)} \subset \mathcal{H}_{\omega_P}$ spanned by the vectors
$$ \big\{ Z\, \Omega_{P} \vert\, Z \in \mathcal{V}_{\omega_P}^{(\varepsilon)}\big\}\,.$$
By $Q^{(\varepsilon)}$ we denote the orthogonal projection onto $\mathcal{W}_{\omega_P}^{(\varepsilon)}$.

Let $\hat{X}\in\mathcal{O}_S$ be a physical quantitiy characteristic of $S$, and let \mbox{$X(P)\in \mathcal{E}_P$} denote the self-adjoint operator representing $\hat{X}$. We consider the spectral decomposition of $X(P)$:
\begin{equation}\label{spect-dec}
X(P)=\sum_{k=1}^{M} x_{j}\, \Pi_{j}(P)\,, 
\end{equation}
where the operators $\Pi_{k}(P) \in \mathcal{E}_{P}, k=1,...,M\leq \infty,$ are the spectral projections of $X(P)$, with 
$$\Pi_{k}(P) = \Pi_{k}(P)^{*}\,,\,\Pi_{j}(P)\cdot \Pi_{k}(P) =\delta_{jk}\, \Pi_{j}(P), \,\forall j, k\,, \,\,\sum_{k=1}^{M} \Pi_{k}(P) = {\bf{1}},$$ and $x_1, ..., x_M$ are the eigenvalues of $X(P)$ ($=$ eigenvalues of $\hat{X}$), ordered in such a way that the sequence 
$\big(\omega_{P}(\Pi_{k}(P))\big)_{k=1}^{M}$ is \textit{decreasing}. 
Let $L\leq M$ be such that
$$\omega_{P}({\bf{1}}- \sum_{k=1}^{L} \Pi_{k})< \varepsilon\,.$$
Given an operator $A \in \mathcal{E}_P$, we denote by $\epsilon_{\omega_P} (A)$ the \textit{unique} operator in the algebra 
$\mathcal{V}_{\omega_P}^{(\varepsilon)} \subset \mathcal{Z}_{\omega_P}$ given by
\begin{equation}\label{cond-exp}
Q^{(\varepsilon)} A \Omega_{P}=: \epsilon_{\omega_P}(A) \Omega_P\,, \qquad \epsilon_{\omega_P}(A) 
\in \mathcal{V}_{\omega_P}^{(\varepsilon)}\,.
\end{equation}
The map 
$$\epsilon_{\omega_P}: \mathcal{E}_{P} \rightarrow \mathcal{V}_{\omega_P}^{(\varepsilon)}$$ 
is called a \textit{``conditional expectation''}; (see \cite{Takesaki} for a systematic theory). 
Claiming that a measurement of the physical quantity $\hat{X}$ can be expected to be possible and to record the event 
\mbox{$\big\{\pi_{\xi} \vert\, \xi \in \textfrak{X} \big\}$} generating 
$\mathcal{Z}_{\omega_P}$ with a resolution of order $\varepsilon$ relies on the validity of the following\\

\underline{Basic Assumption}:
\begin{equation}\label{recording-event}
\Vert \Pi_{k}(P) - \epsilon_{\omega_P}\big(\Pi_{k}(P)\big)\Vert < \varepsilon, \quad \forall \, k=1,..., L\,.
\end{equation}
It is not hard to verify (but see \cite{Fr2}, Eqs. (22), (23), for a proof) that this Assumption implies that
\begin{equation}\label{mixture}
\omega_{P}(A)= \sum_{k=1}^{L} \omega\big(\Pi_{k}(P) \,A \, \Pi_{k}(P)\big) + \mathcal{O}\big(L\,\varepsilon\, \Vert A \Vert \big)\,,\qquad \forall \, A \in \mathcal{E}_{P}\,,
\end{equation}
i.e., the state $\omega_{P}$ is an {\bf{incoherent}} superposition of eigenstates of the operator $X(P)$, up to an error 
of order $\varepsilon$. In this very precise sense, one can say that Assumption \eqref{recording-event} implies that there is an approximate measurement of the physical quantity $\hat{X}$ in the future of the point $P$.

Using a simple lemma (see \cite{FS-Vienna}, Lemma 8 and Appendix C), one can show that if $\varepsilon$ is sufficiently small Assumption \eqref{recording-event} implies that there are orthogonal projections $\pi_{k}(\hat{X}) \in \mathcal{Z}_{\omega_P}$ with the property that 
$$\Vert \Pi_{k}(P) - \pi_{k}(\hat{X})\Vert < \mathcal{O}(\varepsilon),$$
and
$$\omega_{P}(A)= \sum_{k=1}^{L} \omega\big(\pi_{k}(\hat{X}) \,A \, \pi_{k}(\hat{X})\big) + 
\mathcal{O}\big(L\,\varepsilon\, \Vert A \Vert \big)\,,\qquad \forall \, A \in \mathcal{E}_{P}\,.$$
In this precise sense, if $L\geq 2$ a measurement of the quantity $\hat{X}$ in the future of $P$ yields non-trivial information about the {\bf{event}} described by $\mathcal{Z}_{\omega_P}$ happening in the future of $P$. If $L=N$ the projections $\big\{\pi_{k}(\hat{X})\vert k=1,...,L\big\}$ must coincide with the projections $\big\{\pi_{j}\vert j=1,...,N\big\}$ introduced right before \eqref{lower-bound}, provided $\varepsilon \ll 1$ is sufficiently small. In this case, a measurement of 
$\hat{X}$ yields very precise information about the event happening in the future \mbox{of $P$.}

For further discussion of these matters see \cite{Fr2}, (Sect. 3, V.).

\section{Conclusions and outlook}

In this last section, some scattered remarks and speculations that grow out of the results sketched in Sections 4 and 5 are presented.

\begin{enumerate}
\item{In our attempt to cast \textit{local relativistic quantum theory} in a form compatible with the manifestation of what we have defined to be \textit{``events''} and with a solution of the \textit{``measurement problem''}, the \textit{``Principle of Diminishing
 Potentialities'' (PDP)}, (see Definition 3, Sect.~4, Eq.~\eqref{PDP}, and \cite{Fr2}), plays a fundamental role. We have seen that if space-time is \textit{even-dimensional} (e.g.,
$\mathcal{M}=\mathbb{M}^{4}$) and if there exist massless \mbox{particles --} photons, gravitons and, possibly, Dark-Energy modes -- satisfying some form of Huygens' Prinicple, (see \cite{Buch1}), then $(PDP)$ holds. One may argue that $(PDP)$ also 
holds in space-times containing blackholes. From a very general point of view, it appears that a quantum theory satisfying 
$(PDP)$ is necessarily ``relativistic'', and the dimension of its space-time must be even.}

\item{In Definitions 3 and 4 of Sect. 4, we have seen that there is a purely algebraic way to equip space-time $\mathcal{M}$ with a \textit{causal structure}: A space-time point $P$ is in the past of a space-time point $P'$ (written as $P \prec P'$) iff
$$\mathcal{E}_{P'} \subsetneqq \mathcal{E}_{P}, $$
and the relative commutant, $\mathcal{E}'_{P'} \cap \mathcal{E}_P$, of the algebra $\mathcal{E}_{P'}$ in $\mathcal{E}_{P}$ is a non-commutative algebra.
Two points $P$ and $P'$ are space-like separated (written as $P\bigtimes P'$) iff $P$ is not in the past of $P'$ and $P'$ is not in the past of $P$. It would be desirable to further elucidate the relationship of the algebras $\mathcal{E}_P$ and $\mathcal{E}_{P'}$ in case the points $P$ and $P'$ are space-like separated.

Ultimately, we would like to {\bf{reconstruct}} space-time from purely algebraic data concerning a family (or families) of operator algebras equipped with certain relations, in particular inclusions and statements about relative commutants, given a state on these algebras. A (presumably not entirely successful) attempt in this direction has been made in \cite{Bannier}.
}
\item{In the formalism described in Sect.4, {\bf{``events''}} are localized in the future of certain space-time points, $P$; in the sense that they are described in terms of the abelian algebras $\mathcal{Z}_{\omega_P} \subset \mathcal{E}_P$, where, for a given point $P$, $\mathcal{Z}_{\omega_P}$ is the center of the centralizer of the state $\omega_P$ on the algebra $\mathcal{E}_P$, with $\mathcal{E}_P$ describing all potentialities in the \textit{future} of $P$. The \textit{actual event} happening in the future of some point $P$ is an orthogonal projection, 
$\pi^{P}_{\xi}$, belonging to $\mathcal{Z}_{\omega_P}$, for some point $\xi$ in an index set $\textfrak{X}^{P}$, and having a strictly positive probability as predicted by \textit{Born's Rule}. In view of \mbox{Axiom 2,} Sect.~4, it would be important to have a more precise idea about the space-time regions where the operators $\pi^{P}_{\xi}, \xi \in \textfrak{X}^{P}$, are localized. This might actually yield information about the {\bf{geometry}} of space-time and, ultimately, support the view that gravitation 
is an ``emergent'' (or ``derived'') phenomenon.

To render these remarks a little more precise, we recall that one expects that all the algebras $\mathcal{E}_P$ 
are isomorphic to a ``universal'' 
von Neumann algebra $\textfrak{N}$. One would like to know more about properties of states, $\omega$, on $\textfrak{N}$ for which the centers, $\mathcal{Z}_{\omega}(\textfrak{N})$, of the centralizers $\mathcal{C}_{\omega}(\textfrak{N})$ of $\omega$ are non-trivial, in the sense of \mbox{Definition 7,} Sect.~4. In \cite{Fr2}, 
\begin{equation}\label{NC-spectrum}
\textfrak{Z}_{S}:= \bigcup_{\omega} \mathcal{Z}_{\omega}(\textfrak{N}),
\end{equation}
where $\omega$ ranges over all ``states of physical interest'', has been dubbed the \textit{``non-commutative spectrum''} of the system $S$. It is the ``state space'' of the stochastic branching process defined by Eqs.~\eqref{History}, \eqref{state-propagation} and \eqref{hist-prob} of Sect.~4, which describes the \textit{stochastic evolution of states} of $S$. Unfortunately, we have very little insight into the structure of the non-commutative spectrum $\textfrak{Z}_{S}$. 

It would be important to equip the algebra $\textfrak{N}$ (and hence $\mathcal{E}_P$, for \mbox{$P\in \mathcal{M}$)}
with a local structure, (in the sense that $\textfrak{N}$ is generated by a net of local sub-algebras),
and to attempt to show that \textit{events},
i.e., elements of one of the algebras $\mathcal{Z}_{\omega}(\textfrak{N})$, with $\omega$ a ``state of physical interest'', \textit{are typically contained in sub-algebras of 
$\textfrak{N}$ corresponding to what can be considered a ``bounded region'' of space-time}. This would help to introduce a more precise version of Axiom 2. But this topic, too, remains to be clarified.
}
\item{One would expect that, for initial conditions given by states, $\omega_{\Sigma_0}$, of $S$ of ``physical interest'', 
(see Eq.~\eqref{algebra}, Sect.~4), the ensemble of events happening in the future of the points belonging to a foliation 
$\big\{\Sigma_{\tau}\vert \tau \in [0,1]\big\}$ of some slab of space-time (see Sect. 4, after Axiom 2) is \textit{countable}, and that these events are localizable in bounded regions of space-time. One would expect, moreover, that the metric extension of a space-time region within which an event can be localized is constrained by \textit{space-time uncertainty relations} of a kind discussed, e.g., in \cite{Fredenhagen}. This ought to be a consequence of time-energy uncertainty relations and of the possibility  that blackholes form in the aftermath of energetic events, which, afterwards, would evaporate.

Alas, I don't know how to even start to derive these expectations from a more precise formalism of local relativistic quantum theory. Yet, the results reviewed in this paper and in \cite{Les-Diablerets} suggest that, once we truly understand what is meant by a local relativistic quantum theory of events, we will view {\bf{events}} as the {\bf{basic building blocks}} weaving the fabric of space-time and the \textit{relations between events} as determining the {\bf{geometry of space-time}}.
}
\end{enumerate}
To conclude, I want to express the hope that the results, problems and speculations reviewed in this paper might 
challenge colleagues with more technical knowledge and strength than I am able to muster to go further towards 
the goal of truly understanding the miracles of quantum theory.


\begin{thebibliography}{References}

\bibitem{Carroll}Sean Carroll, in the 'New York Times'

\bibitem{BFS-forks} Ph. Blanchard, J. Fr\"{o}hlich and B. Schubnel, \textit{A 'Garden of Forking Paths' -- the Quantum Mechanics of Histories of Events}, Nucl. Phys. B{{\bf912}} (2016), 463-484

\bibitem{Fr2} J. Fr\"ohlich, \textit{A brief review of the $ETH$ Approach to Quantum Mechanics}, to appear in: “Frontiers in Analysis and Probability”, N. Anantharaman and A. Nikeghbali (eds.), Springer-Verlag (2020) [arXiv:1905.06603]

\bibitem{Haag} R. Haag, \textit{Fundamental Irreversibility and the Concept of Events}, Commun. Math. Phys. {\bf{132}} (1990), 245-251;\\
R. Haag, \textit{Events, Histories, Irreversibility}, in: \textit{Quantum Control and Measurement}, Proc. ISQM, ARL Hitachi, H. Ezawa and Y. Murayama (eds.), North Holland, Amsterdam 1993;\\
Ph. Blanchard and A. Jadczyk, \textit{Event-Enhanced Quantum Theory and Piecewise Deterministic Dynamics}, Annalen der Physik {\bf{4}} (1995), 583-599

\bibitem{Kay} Bernard S. Kay and Varqa Abyaneh, \textit{Expectation values, experimental predictions, events and entropy in quantum gravitationally decohered quantum mechanics}, arXiv:0710.0992 (v1), unpublished;\\
Bernard S. Kay, \textit{The Matter-Gravity Entanglement Hypothesis}, Found Phys {\bf{48}} (2018), 542-557

\bibitem{GRW} G. C. Ghirardi, A. Rimini, and T. Weber, \textit{Unified dynamics for microscopic and macroscopic systems}, Phys. Rev. D {\bf{34}} (1986), 470

\bibitem{Wigner} E. P. Wigner, \textit{Remarks on the mind-body question}, in: I. J. Good, "The Scientist Speculates", Heinemann, London 1961

\bibitem{Hardy} L. Hardy, \textit{Quantum mechanics, local realistic theories, and Lorentz-invariant realistic theories}, Phys. Rev. Letters. {\bf{68}}, (20) (1992), 2981–2984; and\\
\textit{Nonlocality for two particles without inequalities for almost all entangled states}, Phys. Rev. Letters. {\bf{71}}, (11) (1993), 1665–1668. 

\bibitem{Renner} D. Frauchiger and R. Renner, \textit{Quantum Theory Cannot Consistently Describe the Use of Itself}, Nature Communications {\bf{9}} (2018), \# 3711 

\bibitem{Bell-book} J. S. Bell, \textit{Speakable and Unspeakable in Quantum Mechanics}, Cambridge University Press, Cambridge (UK) 1987.\\
See also:\\
J. A. Wheeler and W. H. Zurek, \textit{Quantum Theory and Measurement}, Princeton University Press, Princeton NJ, 1983;\\
K. Hepp, \textit{Quantum Theory of Measurement and Macroscopic Observables}, Helv. Phys. Acta {\bf{45}} (1972), 237-248;\\
H. Primas, \textit{Asymptotically Disjoint Quantum States}, in: \textit{Decoherence: Theoretical, experimental and Conceptual Problems}, pp 161-178, Ph. Blanchard, D. Giulini, E. Joos, C. Kiefer and I.-O. Stamatescu (eds.), Springer-Verlag, Berlin 2000

\bibitem{FFS} J. Faupin, J. Fr\"{o}hlich and B. Schubnel, \textit{On the Probabilistic Nature of Quantum Mechanics and the Notion of 'Closed' Systems}, Ann. Henri Poincar\'e {\bf{17}} (2016), 689-731

\bibitem{FS-Prob-Th} J. Fr\"{o}hlich and B. Schubnel, \textit{Quantum Probability Theory and the Foundations of Quantum Mechanics}, arXiv:1310.1484, in: \textit{The Message of Quantum Science -- Attempts Towards a Synthesis}, Ph. Blanchard and J. Fr\"{o}hlich (eds.), Springer-Verlag, Berlin-Heidelberg-New York 2015


\bibitem{Fr1} J. Fr\"{o}hlich, \textit{Quantum Theory and Causality}, Talks at the University of Leipzig (2018), TU-Stuttgart (2019), IHES (2019) and at Vietri sul Mare (Italy) (2019) -- slides available on `ResearchGate'.

\bibitem{Durr-Teufel} D. D\"{u}rr and S. Teufel, \textit{Bohmian Mechanics -- The Physics and Mathematics of Quantum Theory}, Springer-Verlag, Berlin-Heidelberg-New York 2009

\bibitem{Jost} R. Jost, \textit{The General Theory of Quantized Fields}, AMS Publ., Providence R.I., 1965.

\bibitem{Brat-Rob} O. Bratteli and D. W. Robinson, \textit{Operator Algebras and Quantum Statistical Mechanics, Vol. 1}, 
2${nd}$ edition, Springer-Verlag, Berlin-Heidelberg-New York, 1997.

\bibitem{Haag1} R. Haag, \textit{Local Quantum Physics -- Fields, Particles, Algebras}, Springer-Verlag, Berlin-Heidelberg-New York, 1992.

\bibitem{Buch1} D. Buchholz, \textit{Collision Theory for Massless Bosons}, Commun. Math. Phys. {\bf{52}}, (1977), 147-173; (see Theorem 9)

\bibitem{BuDo} D. Buchholz and S. Doplicher, \textit{Exotic infrared representations of interacting systems}, Ann. lnst. H.Poincar\'{e} (Physique th\'{e}orique) {\bf{40}}, no. 2 (1984), 175-184

\bibitem{BR} D. Buchholz and J. E. Roberts, \textit{New Light on Infrared Problems: Sectors, Statistics, Symmetries and Spectrum}, Commun. Math. Phys. {\bf{330}} (2014), 935-972

\bibitem{FS-state-prep} J. Fr\"{o}hlich and B. Schubnel, \textit{The Preparation of States in Quantum Mechanics}, J. Math. Phys. {\bf{57}} (2016), 042 101

\bibitem{FS-Vienna}J. Fr\"{o}hlich and B. Schubnel, \textit{Do We Understand Quantum Mechanics -- Finally?}, arXiv:1203.3678, in: Proceedings of conference in memory of Erwin Schr\"{o}dinger, Vienna, January 2011, publi. in 2012

\bibitem{Schubnel-thesis} B. Schubnel, \textit{Mathematical Results on the Foundations of Quantum Mechanics}, PhD thesis 2014, available at https://doi.org/10.3929/ethz-a-010428944

\bibitem{Les-Diablerets} J. Fr\"{o}hlich, \textit{ 'ETH' in Quantum Mechanics}, Notes of Lectures on the Foundations of Quantum Mechanics, Les Diablerets, January 9 - January 14, 2017

\bibitem{Takesaki} M. Takesaki, \textit{Conditional Expectations in von Neumann Algebras}, J. Funct. Anal. {\bf{9}} (1972), 306-321;\\
F. Combes, \textit{Poids et Esp\'erances Conditionnelles dans les Alg\`ebres de von Neumann}, Bull. Soc. Math. France {\bf{99}} (1971), 73-112

\bibitem{Bannier} U. Bannier, \textit{Intrinsic Algebraic Characterization of Space-Time Structure}, Intl. J. of Theor. Physics {\bf{33}} (1994), 1797-1809

\bibitem{Fredenhagen} S. Doplicher, K. Fredenhagen and J. E. Roberts, \textit{Spacetime quantization induced by classical gravity}, Physics Letters {\bf{331}} (1994), 39-44\\

\end{thebibliography}
\end{document}